\documentclass[intlimits,twoside,a4paper]{article}
\usepackage[cp1251]{inputenc}

\usepackage[eqsecnum]{cmpj3}

\newcommand{\vvec}[1]{\mathrm{\bf #1}}
\newcommand{\vhat}[1]{\hat{\mathrm{\bf #1}}}
\newcommand{\ang}{\mbox{\normalfont\AA}}

\issue{2022}{25}{3}{33601}
\doinumber{10.5488/CMP.25.33601}

\title[Photo-switchable liquid crystalline brush]%
{Photo-switchable liquid crystalline brush as an aligning surface for liquid crystals: modelling via mesoscopic computer simulations}%

\author[D. Yaremchuk,  T. Patsahan, J. Ilnytskyi]{D. Yaremchuk\orcid{0000-0003-2888-5878},
T. Patsahan\orcid{0000-0002-7870-2219}, J.  Ilnytskyi\orcid{0000-0002-1868-5648}
\thanks{Corresponding author: \email{iln@icmp.lviv.ua}.}}
\address{
 Institute for Condensed Matter Physics of the National Academy of Sciences of Ukraine,\\
1 Svientsitskii St., 79011 Lviv, Ukraine
}

\date{Received March 19, 2022, in final form May 16, 2022}

\begin{document}

\maketitle

\begin{abstract}

We consider the mesoscopic model for the liquid crystalline brush that might serve as a photoswitchable aligning surface for preorientation of low molecular weight liquid crystals in a bulk. The brush is built by grafting the polymer chains of a side-chain molecular architecture, with the side chains terminated by a chromophore unit mimicking the azobenzene unit, to a substrate. When irradiated with ultraviolet light, the chromophores photoisomerize into a non-mesogenic \textit{cis} state and the whole system turns into an ordinary polymer brush with no orientational order and two states: the collapsed and straightened one, depending on the grafting density. When irradiated with visible light, the chromophores photoisomerize into a mesogenic \textit{trans} state, resulting in formation of a transient network between chains because of a strong attraction between chromophores. Spontaneous self-assembly of the brush in these conditions results in an orientationally isotropic polydomain structure. The desired uniaxial planar ordering of chromophores within a brush can be achieved at certain temperature and grafting density intervals, as the result of a two-stage preparation protocol. An external stimulus orients chromophores uniaxially at the first stage. The system is equilibrated at the second stage at a given temperature and with the external stimulus switched off. The preoriented chromophores either keep or loose their orientations depending on the strength of the memory effect inherent to a transient network of chains that are formed during the first stage, similarly to the case of the liquid crystalline elastomers, where such effects are caused by the covalent crosslinks.

\printkeywords

\end{abstract}

\section{\label{I}Introduction}

Polymer brush, typically a flat surface containing grafted polymer chains, provides a good example of a functional surface \cite{Zhao2000,Minko2006,Ma2019,Koch2020}. All the factors, such as: grafting density, chemical structure and molecular architecture of polymer chains, as well as the presence of specific responsive groups, play a crucial role in their functionality. One of the examples of such groups are mesogens, that are capable of forming orientationally and spatially ordered liquid crystalline (LC) phases \cite{deGennes1993}. Polymer chains serve as a soft embedding matrix for the mesogens resulting in stabilization or disruption of their LC phases, similarly to the case of LC elastomers \cite{Ula2018}. On the other hand, the presence of mesogens turns this composite, i.e., a liquid crystalline polymer brush (LCPB), into a light-sensitive functional surface. 

Both synthetic protocols \cite{Peng1999} and theoretical treatment \cite{Williams1993,Amoskov1998,Birshtein2000} of LCPBs were started at least two decades ago. Their principal features have  narrowed up to the formation the LC phases with the possibility to achieve either homeotropic or uniaxial planar (UPL) orientation of the mesogens with respect to the substrate. The latter case is thought to be especially useful for practical application in the LC displays technology, where such LCPB can serve as a commanding surface for precise aligning of the bulk LC in a LC cell, see, e.g., \cite{Camorani2009}. Other applications are based on spontaneous tilting transition in the main chain LCPB \cite{Blaber2019}; LC ordering as the result of dewetting of the LCPB \cite{Mukai2019}; the synergy between the LCPB and nanotubes allowing to fabricate smart windows with their transparency controlled by the near infrared light \cite{Deng2021} and so on. 

To introduce more control over the alignement of mesogens in the LCPB, the mesogens in a form of the azobenzene chromophores can be used. By illumination of such chromophores by the light with suitable wavelength, intensity and polarization, their \textit{trans}-isomers are capable of reorienting theirselves perpendicularly to the polarization vector because of their angular-dependent absorption of photons, the so-called ``angular hole burning'' \cite{Todorov1984}. This leads to the possibility of a precise control for the nematic director in the LC phases of the azobenzene containing LCPBs \cite{Uekusa2007,Uekusa2008,Mukai2016,Nagano2018,Li2018},  
photocontrolled nanopatterning~\cite{Seki2009,Nandivada2010,Koskela2014,Kollarigowda2017}, surface relief formation \cite{Lomadze2011,Kopyshev2015}, designing superadhesive surfaces \cite{Roling2016}, making waves in such brushes \cite{Gelebart2017,Jelken2019}, developing smart biointerfaces \cite{Wei2017} and to other applications \cite{Santer2017}.

We  restrict our focus here to the application of the LCPBs as the photocontrolled commanding surfaces for alignment of bulk LCs. After an early work \cite{Ichimura1988} involving the azobenzene-treated substrate, the later research was concentrated on the LCPBs comprising polymers of the side-chain architecture, where the side-chains were terminated by an azobenzene unit. As first demonstrated in \cite{Uekusa2007}, such LCPB organizes into a smectic phase characterized by the planar orientation of azobenzenes,  contrary to the spin-casted film, where the homeotropic orientation prevails. The 2D in-plane orientation of azobenzenes is characterized, in general, by a polydomain structure. It can be transformed into a monodomain sample once the system is illuminated with the linearly polarized beam with its incidence perpendicular to the substrate \cite{Uekusa2008,Mukai2016}. The nematic director in this case is perpendicular to the polarization vector of a beam. This level of the in-plane dichroism is achieved at the temperatures slightly above the glass transition temperature and is not attained for the spin-cast prepared films. In another study \cite{Haque2013}, the azobenzene-containing fragment was grafted to the substrate via linear spacer, and, upon the increase of its length, ``the smectic layer structure  also became assembled more intrinsically with the cooperative wiggling ascribed to the flexible buffer layer'' \cite{Haque2013}. Yet other studies, based on the block-copolymer LCPB, showed that the surface segregation of the low surface free energy block also leads to the planar orientation for the azobenzenes \cite{Nagano2016,Nagano2018}. A gradual alignment of bulk mesogens from homeotropic to planar can be achieved by changing the grafting density of the azobenzene-containing LCPB, using the interplay between the preferentially perpendicular orientation of azobenzenes and the backbone and the stretching director of the backbone, controlled by the grafting density \cite{Li2018}.

Summarizing these experimental findings, it is quite evident that in order to apply the LCPB as commanding surfaces for the alignement of bulk mesogens, their orientational order should be (a) controllable, and (b) stable. Requirement (a) can be satisfied by incorporating the azobenzene groups into its molecular architecture and providing their capability to photoisomerize. Requirement (b) needs deeper insight towards both formation and stability of the ordered LC phases, that are inherent for the LCPB of a given molecular architecture at particular grafting density and temperature. These issues are addressed in the current study using mesoscale computer simulations. In this respect, it continues a series of our previous works on self-assembly of decorated nanoparticles \cite{Ilnytskyi2010,Ilnytskyi2013,Slyusarchuk2014,Ilnytskyi2016a,Ilnytskyi2016b} and of their adsorption on the LCPB \cite{Slyusarchuk2020}, by using a similar type of modelling.

The outline of a study is as follows: section \ref{II} contains the description of the model,  section \ref{III} covers the properties of the model azobenzene-containing LCPB under ultraviolet light, section \ref{IV} addresses such a brush under visible light, conclusions are given in section \ref{V}.

\section{\label{II}Mesoscopic model for the photo-switchable liquid crystralline brush}

%
\begin{figure}[!t]
\begin{center}
\includegraphics[clip,width=11cm]{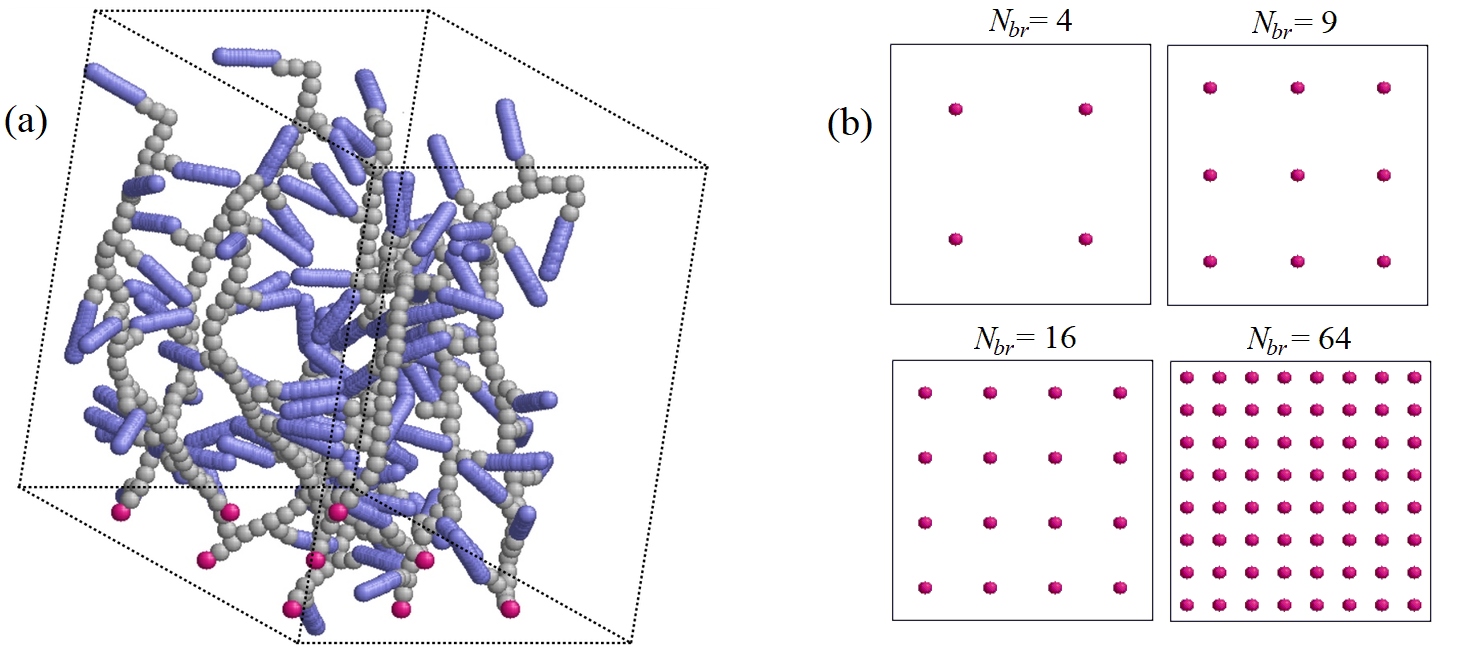}
\caption{\label{BR_model}(Colour online) (a) Snapshot showing the model LCPB of $N_\text{br}=9$ molecules grafted to the substrate. Polymer beads are shown in gray, chromophores --- blue, solvent beads not shown. Monomers grafted to the substrate are shown in pink. (b) Regular patterns of beads grafted on a substrate shown for a selected number of brush molecules $N_\text{br}$ as indicated in the figure.}
\end{center}
\end{figure}

The model LCPB considered here, comprises a set of $N_\text{br}$ polymer molecules of the side-chain architecture, grafted to the bottom surface of the simulation box by the first monomer of each molecule, shown in pink in figure~\ref{BR_model}(a). These grafted monomers are considered to be frozen: their positions are arranged using a regular 2D pattern and are fixed throughout the simulations, see figure~\ref{BR_model}(b). We expect the exact type of a pattern to have a minimal influence on the obtained results as long as it is kept homogeneous in space. Following our previous works \cite{Ilnytskyi2010,Ilnytskyi2013,Slyusarchuk2014,Ilnytskyi2016a,Ilnytskyi2016b,Slyusarchuk2020}, we employ a coarse-grained modelling approach. It utilizes splitting each polymer molecule into a set of connected beads, each representing a specific atomic group, e.g., a fragment of three CH${_2}$ units, ester group, the azobenzene unit, etc. As a result, each molecule contains a backbone of $36$ spherical beads and $12$ side chains, where each side chain contains a spacer of two spherical beads and a terminal spherocylinder bead mimicking the azobenzene chromophore. Because of the coarse-grained nature of a model, the same spherocylinder shape is used for both \textit{trans-} and \textit{cis}-isomers of azobenzene, whereas the differences between them are reflected in different interaction potentials involving these isomers. We consider LCPB in an explicit solvent, represented by spherical beads. To simplify the following notations, we denote spherical beads of a polymer and of a solvent via subscripts $p$ and $s$, respectively; spherocylinder bead representing \textit{trans-} or \textit{cis-}isomer via subscripts $t$ and $c$, respectively, whereas the case of any of these isomers is denoted by $a$. All spherical beads are chosen to have the same diameter of $\sigma_p=\sigma_s=4.59$~{\AA}, spherocylinder chromophore beads are characterised by the diameter $\sigma_a=3.74$~{\AA} of their spherical cap and by the elongation of $L/\sigma_a=3$, where $L$ is the separation between the centers of their spherical caps.

The expression for the intramolecular bonded interactions 
\begin{equation}\label{Vbonded}
\begin{array}{cc}
V^{\mathrm{BON}}=& 
  \displaystyle
  \sum_{i=1}^{m_b}k_b(l_i-l_0)^2 +  \sum_{i=1}^{m_a}k_a(\theta_{i}-\theta_0)^2 + \sum_{i=1}^{m_z}k_z(\zeta_{i}-\zeta_0)^2,
\end{array}
\end{equation}
involves $m_b$ harmonic bonds and $m_a+m_z$ pseudo-valent angle bending terms that control stiffness of the side chains, as it is typically done at this level of coarse-graining \cite{Milano2005,Anderson2017,Lavagnini2020,Lavagnini2021}. The backbone contains no bond bending terms but is semiflexible implicitly, via the presence of the side chains. The values of $l_0=3.617$~{\AA} and $8.59$~{\AA} are used as the refernce bonds for the sphere-sphere and sphere-spherocylinder pairs. The reference angle $\theta_0$ is equal to $\piup$ for the linear fragment of a side chain, and is equal to $\piup/2$ where the attachment of the side chain to the backbone occurs (the triplet includes a branching bead of a backbone, the backbone bead that precedes it, and a first bead of a side chain), to enforce their mutually perpendicular arrangement. The last term in equation~(\ref{Vbonded}) involves the angle $\zeta_0=0$, which ensures collinearity of the terminal spherocylinder to the bond which links its center with the last spherical of a side chain, see \cite{Wilson1993}. For the sake of brevity, we introduce the following energy unit: $\textrm{ j}=10^{-20}\textrm{ J}$. Force constants, expressed via this unit are: of $k_b=50\textrm{ j}/\ang^2$ and $k_a=k_z=20\textrm{ j}/\mathrm{rad}^2$.

Let us consider the nonbonded interactions now. Following the ideology of coarse-grained models (e.g., these employed in the dissipative particle dynamics simulations \cite{GrootMadden1998}), most of them are assumed to be short-ranged soft repulsive, with the strength of the repulsion reflecting the chemical nature of each bead. We assume the polymer $p$ beads and the $t$ chromophores to be non-polar, whereas the solvent $s$ beads and the $c$ chromophores are polar (dipole moment of \textit{trans-} and \textit{cis-}isomers of a typical azobenzene chromophore is discussed in \cite{Akiyama2003}). Higher magnitude for the repulsion parameter is used then to describe the segregation effect between the polar and non-polar beads, see, e.g., \cite{GrootMadden1998}. The same principle can be used for the $\{i,j\}$ pair of interacting spherocylinders by defining their ``internal core'' (the line joining the centers of its two spherical caps) and introducing the minimal separation $d(\vvec{q}_{ij})$ between their cores. Here, $\vvec{q}_{ij}=\{\vhat{e}_i,\vhat{e}_j,\vvec{r}_{ij}\}$ contains all characteristics of their mutual arrangement: the vector $\vvec{r}_{ij}$ between their centers and their respective orientations $\vhat{e}_i$, $\vhat{e}_j$ in space. Then, the same expression for the soft repulsive potential can be used as for the case of spherical beads using minimal separation $d(\vvec{q}_{ij})$ in place of the center-center distance \cite{Kihara1963}. However, it turned out that the soft repulsive spherocylinders form LC phases only in the regime of high aspect ratio \cite{Hughes2005,Hughes2008}. By adding the attractive part to the interactive potential, which represents the dispersion forces, one strengthens the mesogenity of soft spherocylinders and is able to obtain ordered LC phases at much moderate aspect ratios \cite{Lintuvuori2008}. To emphasize a strong attraction between polar $c$ chromophores and a solvent, we opted to use such interaction potential with the attractive contribution for this case, too.

All the cases of the pair interaction potential between $i$th and $j$th beads, discussed above, can be written in a single dimensionless form 
\begin{equation}\label{SAP}
V^{\mathrm{SAP}}[d'(\vvec{q}_{ij})]=\left\{
\begin{array}{ll}
U\big\{[1-d'(\vvec{q}_{ij})]^2-\epsilon'(\vvec{q}_{ij})\big\},& 0\leqslant d'(\vvec{q}_{ij})<1\vspace{2mm},\\
U\big\{[1-d'(\vvec{q}_{ij})]^2-\epsilon'(\vvec{q}_{ij})\big.&\\
\hspace{2em}\big.-\frac{1}{4\epsilon'(\vvec{q}_{ij})}[1-d'(\vvec{q}_{ij})]^4\big\},& 1 \leqslant d'(\vvec{q}_{ij}) \leqslant d'_c\vspace{2mm},\\
0,& d'(\vvec{q}_{ij})>d'_{c},
\end{array}
\right.
\end{equation}
where $U$ defines repulsion strength, $d'(\vvec{q}_{ij})=d(\vvec{q}_{ij})/\sigma_{ij}$ is the dimensionless minimal separation between two beads, and $\sigma_{ij}=(\sigma_{i}+\sigma_{j})/2$ is the length scaling factor for the $\{i,j\}$ interacting pair.
\begin{equation}\label{eps}
\epsilon'(\vvec{q}_{ij})=\Bigg\{4\Big[U'_a-5\epsilon'_1 P_2(\vhat{e}_i\cdot\vhat{e}_j)-5\epsilon'_2\Big(P_2(\vhat{r}_{ij}\cdot\vhat{e}_i)+P_2(\vhat{r}_{ij}\cdot\vhat{e}_j)\Big)
\Big]\Bigg\}^{-1},
\end{equation} 
is a dimensionless well depth of this potential, obtained from the condition that both the expression (\ref{SAP}) and its first derivative on $d'(\vvec{q}_{ij})$ turn to zero when $d'(\vvec{q}_{ij})=d'_{c}$, where $d'_{c}=1+\sqrt{2\epsilon'(\vvec{q}_{ij})}$ is the cutoff separation of the potential. $\vhat{r}_{ij}=\vvec{r}_{ij}/r_{ij}$ is a unit vector connecting the centers of two beads, $U'_a$, $\epsilon'_1$ and $\epsilon'_2$ are dimensionless parameters that define the shape of the interaction potential depending on mutual orientation of beads \cite{Lintuvuori2008}, $P_2(x)=(3x^2-1)/2$ is the second Legendre polynomial.

\begin{figure}[!b]
\begin{center}
\includegraphics[clip,width=7cm]{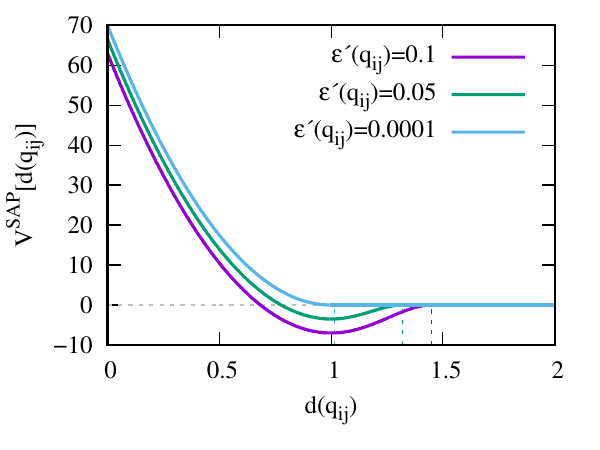}
\caption{\label{BR_potentials}(Colour online) Illustration of the interaction potential $V^{\mathrm{SAP}}[d'(\vvec{q}_{ij})]$ between two beads of any type at $U=70\textrm{ j}$ shown for a range of effective dimensionless well depths $\epsilon'(\vvec{q}_{ij})$ (respective cutoffs $d'_{c}$ are indicated via vertical dashed lines). $d'(\vvec{q}_{ij})$ is the minimum distance between the bead cores, which reduces to the dimensionless separation $r'_{ij}$ between the centers for the case of two interacting spherical beads. At a zero well depth,  $\epsilon'(\vvec{q}_{ij})\to 0$, the potential $V^{\mathrm{SAP}}[d'(\vvec{q}_{ij})]$ turns into a purely repulsive potential $V^{\mathrm{SRP}}[d'(\vvec{q}_{ij})]$, equation~(\ref{SRP}).}
\end{center}
\end{figure}
The effective well depth $\epsilon'(\vvec{q}_{ij})$ influences both the shape of the attractive part and, via $d'_{c}$, its range. When the parameters contained in the expression for $\epsilon'(\vvec{q}_{ij})$, are such that it asymptotically reaches zero, the cutoff $d'_{c}$ approaches $1$ and the interval for the second line in equation~(\ref{SAP}) shrinks to zero. As a result, in this limit, one retrieves the soft repulsive quadratic potential
\begin{equation}
V^{\mathrm{SRP}}[d'(\vvec{q}_{ij})]=\left\{
\begin{array}{ll}
U\left[1-d'(\vvec{q}_{ij})\right]^2, &0\leqslant d'(\vvec{q}_{ij})\leqslant 1\vspace{2mm}\\
0, &d'(\vvec{q}_{ij}) > 1,
\end{array}
\right.
\label{SRP}
\end{equation}
that is used typically in the dissipative particle dynamics simulations \cite{GrootMadden1998}. This limit is illustrated in figure~\ref{BR_potentials} for the case of two spherical beads and the repulsive energy strength of $U=70\textrm{ j}$.

Now we provide  the parameters for the pair interaction between various combinations of beads that are based on their chemical origin. The spherocylinder $tt$ and $cs$ pairs interact via attractive potential~(\ref{SAP}) with $U_{tt}=70\textrm{ j}$, $U'_a=21.43$, $\epsilon'_1=1.714$ and $\epsilon'_2=-1.714$ (the ``model A'' in the nomenclature of~\cite{Lintuvuori2008}). For the spherical $s$ bead we set $\vhat{e}_i=0$. All the rest pair interactions are soft repulsive (\ref{SRP}), where  $U_{pp}=U_{ss}=U_{cc}=U_{ct}=70\textrm{ j}$ and $U_{ps}=U_{ts}=140\textrm{ j}$ to mimic a segregation between polar and non-polar beads. 

The angular hole burning effect, discussed in section \ref{I}, can be modelled either explicitly, via kinetic equations for the fractions of the $t$ and $c$ chromophores \cite{Ilnytskyi2015,Ilnytskyi2019}, or implicitly, via effective field that reorients the $t$ chromophores prefferably perpendicularly to the polarization vector of the incident light  \cite{Ilnytskyi2006,Ilnytskyi2011,Ilnytskyi2016b}. As was shown in \cite{Ilnytskyi2015}, explicit modelling leads to a statistical effect of reorientation of the $t$ chromophores, and this effect can be mapped onto the effective field with the strength dependent on the photoisomerization rates that appear in respective kinetic equations. Here, we choose the simpler approach of two, the implicit modelling of the angular hole burning effect, by introducing the effective field via respective energy term 
\begin{equation}\label{urot}
U^{\mathrm{ROT}} = F\sum_{i=1}^{n_t} P_2(\vhat{e}_i\cdot\vhat{p}),
\end{equation}
where $F>0$ is the effective field strength, $n_t$ is the number of the $t$ chromophores at a given time instance, and unit vectors $\vhat{e}_i$ and $\vhat{p}$ provide the orientation of $i$th $t$ chromophore and of the polarization vector, respectively.

The chromophore is assumed to photoisomerise into the $c$ state under ultraviolet (UV) light and to undergo a reverse photoisomerisation to the $t$ state under visible (Vis) one, following the properties of NO$_2$ substituted azobenzenes \cite{Natansohn1995}. To simplify the description, we consider two limit cases: (i) UV light: all chromophores are in the $c$ state, and (ii) Vis light: all chromophores are in the $t$ state. Simulation box is of a cuboid shape with the following dimensions: $L_x=L_y=100$~{\AA} and $L_z=130$~{\AA} with the periodic boundary conditions applied along the $OX$ and $OY$ axes and with elastic reflective walls at both the $z=0$ and $z=L_z$ planes mimicking a substrate. Polymer molecules are grafted to the bottom plane, $z=0$, with the reduced grafting density $\rho'_{g}=N_\text{br}\sigma_p^2/(L_xL_y)$, expressed via $N_\text{br}$, the interval of $N_\text{br}=1-64$ is considered. Maximum reduced grafting density is, therefore, equal to $0.135$. Simulation box interior is filled with a solvent to keep the total density of about 0.5~$\mathrm{g}/\mathrm{cm}^3$.

Computer simulations are performed by molecular dynamics in the NVT ensemble with the time step  of 20~$\mathrm {fs}$. It is known that such simulations often suffer from covering only the local region of a phase space, especially when the symmetry is spontaneously broken, due to formation of an ordered structure. This shortcoming is, typically, the main source of the uncertainties for the average properties of interest \cite{Grossfield2019} and, therefore, achieving sufficient sampling accuracy for them requires performing several independent simulation runs. The second source of the uncertainties is the finite length of each independent run. To address these requirements, we perform a sequence of $10$ independent runs, each duration of 100~$\mathrm{ns}$,  for each chosen set of system parameters. Each  run of this type consists of a ``reset'' stage, when the LCPB is ``shaken'' at a high temperature to remove the memory on its initial structure, and a ``relaxation'' stage, 
performed at the desired set of parameters. The results at the end of the relaxation stage, during each independed run, are collected to perform averaging of  the properties of interest and estimation of their respective errors. The details regarding the relaxation stages in each particular case are given below. 

In a general sense, the quest of this study, the UPL phase, is an orientationally ordered morphology formed in the LC polymer system incorporating mesogenic $t$-beads. The choice of parameters for the interaction potentials provided above, leads to the occurence of such a phase for similar polymer systems at the temperature $T_c \sim 500$~K \cite{Ilnytskyi2010,Ilnytskyi2013,Slyusarchuk2014,Ilnytskyi2016a,Ilnytskyi2016b}. It was also shown \cite{Ilnytskyi2016b} that the defect-free orientationally ordered phase can be obtained via the self-assembly at the temperatures just below $T_c$. These considerations provide an argument for our choice of working temperatures, $T=480-490$~K, in this study. We also note that molecular systems incorporating non-mesogenic $c$-chromophores do not form orientationally ordered morphologies and are characterized by a smooth dependance of their structure on the temperature.  

\section{\label{III}Photo-switchable liquid crystralline brush under ultraviolet light}

We consider the idealized case, when illumination of the LCPB by the UV light results in photoisomerization of all existent chromophores into the $c$ type. Let us consider different energy terms that affect the structure of LCPB in a solvent.

Each polymer molecule is expected to be of a stretched bended conformation because of the stiffness of their backbones induced by the presence of the side chains. Side chains with terminal chromophores are perpendicular to the backbones due to their strong coupling to them. The molecules may rotate around their grafted beads as semi-rigid objects subjected to the effect of excluded volume with respect to their neighbours. Excluded volume effect is strengthened by solvation of the $c$ chromophores. These properties reduce the LCPB under UV light to the non-LC brush, composed of semi-rigid side chain polymers. The bottom surface of the simulation box has adsorbing properties for the molecules, because the polymer~$p$ beads have less repulsive neighbours near the surface than in a bulk. The competition between the (i) adsorption of the polymers on the botom surface, and (ii) their repulsion because of overcrowding will determine the final state of the LCPB.

We found basically the same structure of the LCPB in a wide temperature range, from $450$~K to $600$~K, therefore, the case of the temperature $480$~K is illustrated only. As discussed above, the simulation performed at each number of molecules $N_\text{br}$ consists of $10$ independent runs. Each run comprises the 10~$\mathrm{ns}$ reset stage performed at high temperature of $T=800$~K followed by the 90~$\mathrm{ns}$ relaxation stage performed at $T=480$~K. Last 20~$\mathrm{ns}$ of the relaxation stage of each independent run serve as a productive part of the simulations and provide statistics for averaging all the properties of interest and their error estimates. For the sake of brevity, thereafter we merge the time evolution of relevant properties of interest obtained in the course of $10$ independent runs into a single plot. Data within productive parts are emphasized via bold lines.

\begin{figure}[!b]
\begin{center}
\includegraphics[clip,width=15cm]{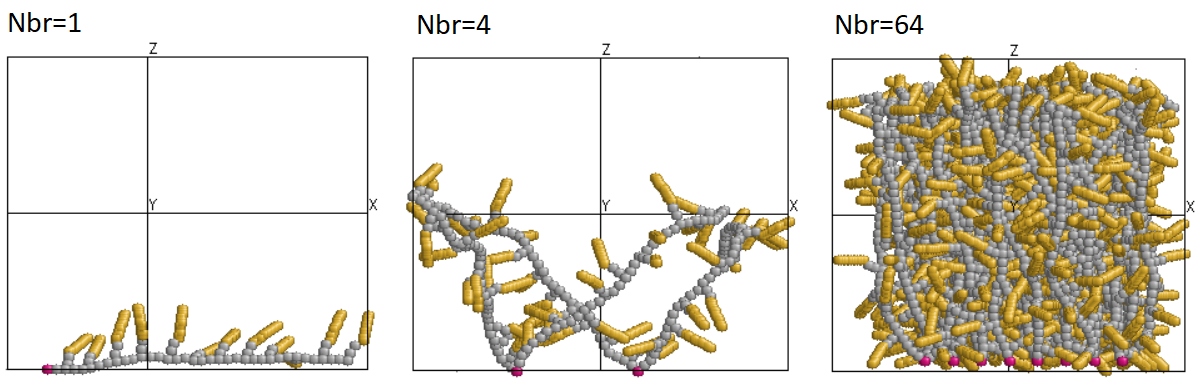}
\caption{\label{BR_snap1}(Colour online) Internal structure of the LCPB under UV illumination at various number of grafted molecules $N_\text{br}$, indicated in the plots. Polymer $p$ beads are coloured in gray, the $c$ chromophores are yellow, solvent beads are not shown. Temperature is $T=480$~K.}
\end{center}
\end{figure}

At small number $N_\text{br}$ of molecules in the LCPB, the effect (i) prevails and we observe a collapsed state, see figure~\ref{BR_snap1}, on the left. With an increase of $N_\text{br}$, the molecules desorb from the bottom surface and straighten up, along the $OZ$ axis, see figure~\ref{BR_snap1}, in the middle and on the right.

The state of the LCBR is characterized by its height, which we normalize with respect to the box dimension $L_z$ along the $OZ$ axis, for the sake of convenience
\begin{equation}
h = \frac{2}{L_z}\frac{\int_{0}^{L_z}z\rho(z)\rd z}{\int_{0}^{L_z}\rho(z)\rd z},
\end{equation}
where $\rho(z)$ is the density profile for the brush beads, obtained in the course of computer simulations. The orientational order of chromophores is fully characterized  by three order parameters, $S_x$, $S_y$ and $S_z$ with respect to the $OX$, $OY$ and $OZ$ spatial axes, defined as 
\begin{equation}
S_{\alpha} = \langle P_2(\vhat{e}_i\cdot\vhat{i}_{\alpha})\rangle,
\end{equation}
where $\vhat{e}_i$ is the unit vector providing spatial orientation of $i$th chromophore, whereas $\vhat{i}_{\alpha}$ is the unit vector directed along the spatial axis indexed via $\alpha=\{x,y,z\}$. The averaging is performed over all chromophores within the LCPB. The substrate is defined in the $XY$ plane at $z=0$ and $z=L_z$. Therefore, the conditions for the UPL structure, relevant to this study, require that $S_z$ is negative with its magnitude essentially different from zero (indicating planarity of chromophores with respect to a substrate), and one of the order parameters $S_x$ or $S_y$ is positive and essentially different from zero (indicating sufficient uniaxial order of chromophores within the $XY$ plane).

\begin{figure}[!t]
\begin{center}
\includegraphics[clip,width=10cm]{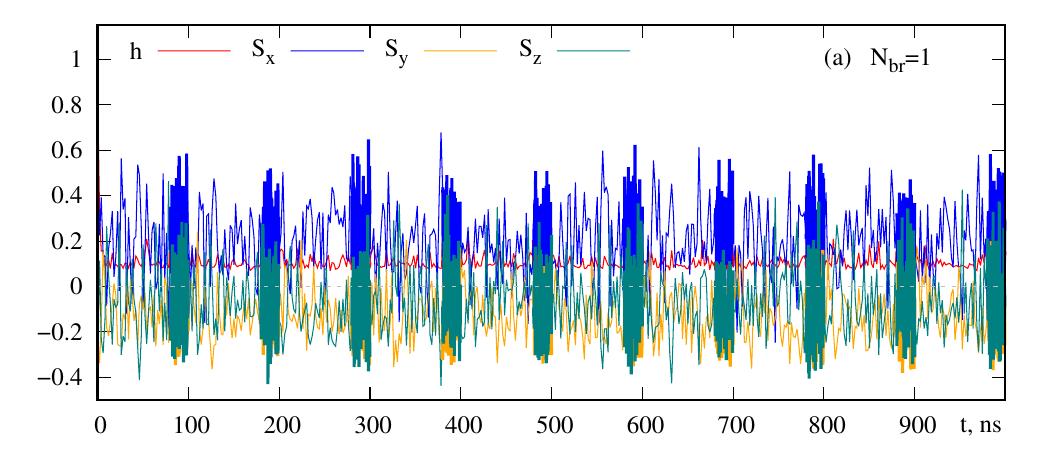}
\includegraphics[clip,width=10cm]{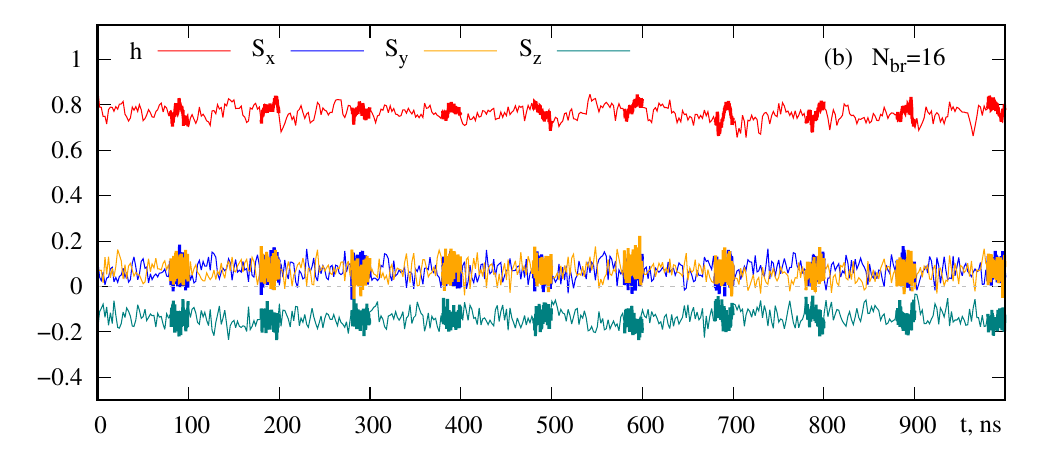}
\caption{\label{BR_uv_evol}(Colour online) Time evolutions of reduced height $h$ and the orientational order parameters $S_{\alpha}$ of the LCPB under UV light shown at (a) $N_\text{br}=1$ and (b) $N_\text{br}=16$, the temperature is $T=480$~K.}
\end{center}
\end{figure}
The time evolutions of all these properties, $h$, $S_x$, $S_y$ and $S_z$, are shown in figure~\ref{BR_uv_evol} for two characteristic cases, $N_\text{br}=1$ and $N_\text{br}=16$. The former is characterizes by low values of $h$ (collapsed brush) and all $S_{\alpha}$ fluctuating close to zero (random orientations). At $N_\text{br}=16$, the brush straightens up ($h\approx 0.8$) and is characterized by weak planarity ($S_z\approx -0.15$) but is not uniaxial (both $S_x$ and $S_y$  fluctuate around $0.05$). Therefore, the transition between the collapsed and straightened states of the LCPB, both orientationally isotropic in the $XY$ plane, occurs somewhere between $N_\text{br}=1$ and $N_\text{br}=16$ in the conditions mimicking the UV light, when most chromophores are in the \textit{cis} state.

\begin{figure}[!t]
\begin{center}
\includegraphics[clip,width=10cm]{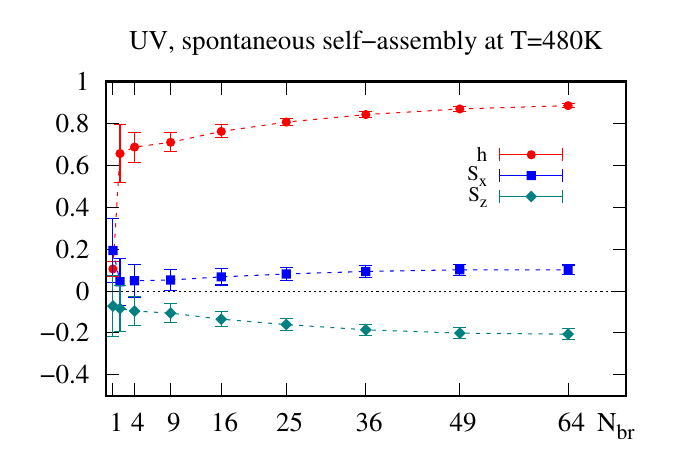}
\caption{\label{BR_Fig_1}(Colour online) Reduced height $h$ and the orientational order parameters $S_z$ and $S_x$ of the LCPB under UV light as the functions of the number of grafted molecules $N_\text{br}$, the temperature is $T=480$~K.}
\end{center}
\end{figure}
This transition is examined  more in detail in figure~\ref{BR_Fig_1}, where the values of $h$, $S_z$ and $S_x$, averaged over productive parts of simulations (shown in bold in figure~\ref{BR_uv_evol}), are shown as the functions of $N_\text{br}$. The results shown indicate that the transition occurs in between $N_\text{br}=1$ and $N_\text{br}=4$. At $N_\text{br}\geqslant 4$, $S_z$ reaches the values about $-0.2$, indicating a certain level of perpendicularity of chromophores to the $OZ$ axis. This is achieved because of the combination of two factors: stretchening of backbones along this axis (excluded volume driven ``dense brush'' regime) and strong coupling of side chains to the backbone. This opens up the possibility of manipulating the orientation order of chromophores by stretching or coiling the backbones, as remarked in the experimental works \cite{Li2018,Blaber2019,Yanagimachi2019}. However, the order parameter $S_x \leqslant 0.1$ for all $N_\text{br}$ being examined (the same holds for the magnitude of $S_y=-S_z-S_x$, not shown), indicating the isotropy of chromophores in the $XY$ plane. On the top of that, the $c$ chromophores are non-mesogenic. Therefore, the LCPB under UV light may be used for inducing only a random planar orientation for low molecular weight LC in bulk. In the next section we  analyse the same set of characteristics for the LCBR under Vis light, where chromophores are strongly mesogenic ones of the $t$ type. 

\section{\label{IV}Photo-switchable liquid crystralline brush under visible light}

We consider here another idealized case, when illumination of the LCPB by the Vis light results in a photoisomerization of all existent chromophores into the $t$ type ones. Within the simulation model, the switch from the $c$ to $t$ type chromophores, changes some of the energy contributions to the free energy. Namely, the pairs of mesogenic $t$ beads interact via the attractive potential (\ref{SAP}) with the parameters provided in section \ref{II}. Additional attraction is attributed to a stronger repulsion between unpolar $t$ and polar $s$ beads, reflected in the higher repulsion strength $U_{ts}=140\textrm{ j}$ as compared to that for the $c$-$s$ pair, see section \ref{II}. Both interactions involving $t$ beads cause the formation of strong $t$-$t$ links, and, as a result, turns the LCPB into a macromolecular network with physical crosslinks \cite{Ilnytskyi2017,Ilnytskyi2018}, with the possibility to form a globally ordered LC phase.

We try different approaches to produce the LCPB network with the properties of an UPL LC phase. The simplest one is to leave the initially disordered system to an unaided self-assembly. The process is characterized by a spontaneous breakage of the symmetry of the LCPB, therefore, we average over $10$ such attempts performed at each $N_\text{ch}$. Each attempt comprises the 10~$\mathrm{ns}$ reset stage performed at $T=800$~K, followed by the 90~$\mathrm{ns}$ unaided self-assembly stage performed at a chosen temperature. Last 20~$\mathrm{ns}$ of the self-assembly stage of each independent run serve as a productive part of the simulations. Following our experience with the mesogens of the same type \cite{Ilnytskyi2010,Ilnytskyi2013}, the self-assembly runs are performed at $T=490$~K, just below the phase transition to the isotropic phase. 

\begin{figure}[!t]
\begin{center}
\includegraphics[clip,width=12cm]{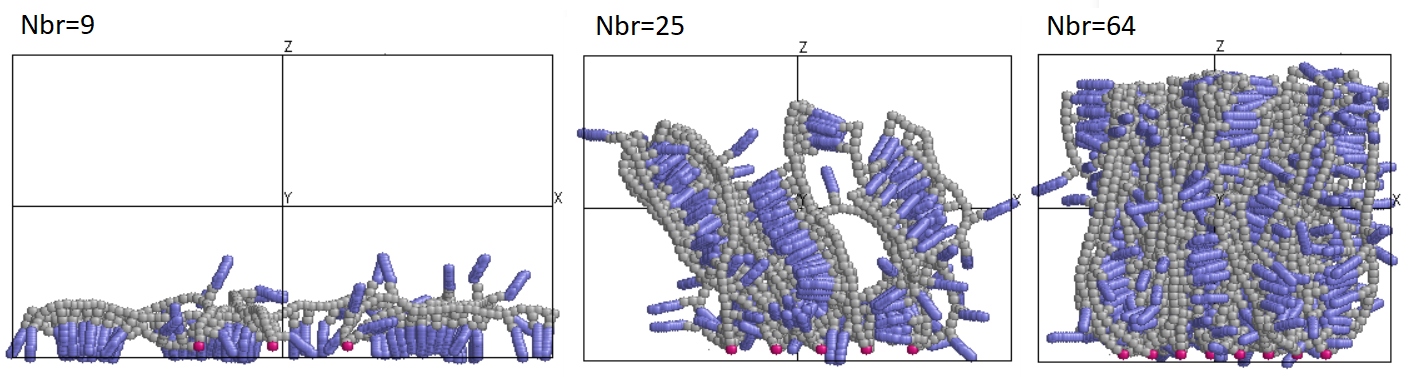}
\caption{\label{BR_snap2}(Colour online) Internal structure of the LCPB under Vis illumination at various number of grafted molecules $N_\text{br}$, indicated in the plots. Polymer $p$ beads are coloured in gray, the $t$ chromophores are in blue, solvent beads are not shown. Unaided self-assembly at $T=490$~K.}
\end{center}
\end{figure}
Some illustrations of the resulting internal structure of the LCPB are shown in figure~\ref{BR_snap2} for various number of grafted molecules $N_\text{br}$. Similarly to the case of UV light, discussed in section \ref{III}, one observes a collapsed state of a brush at small $N_\text{br}$, as a result of adsorption of chains on a substrate (see, left-hand frame in figure~\ref{BR_snap2}). The adsorption is aided by the wall-induced layering of mesogenic \textit{trans} chromophores, as seen in the figure. Stretched state of a LCPB is observed at medium and large $N_\text{br}$ (see, middle and right-hand frames in figure~\ref{BR_snap2}), with strong $t$-$t$ links clearly visible in the figure. In general, the LCPB has a domain-like structure with no prevailing nematic director.   

\begin{figure}[!b]
\begin{center}
	\includegraphics[clip,width=10cm]{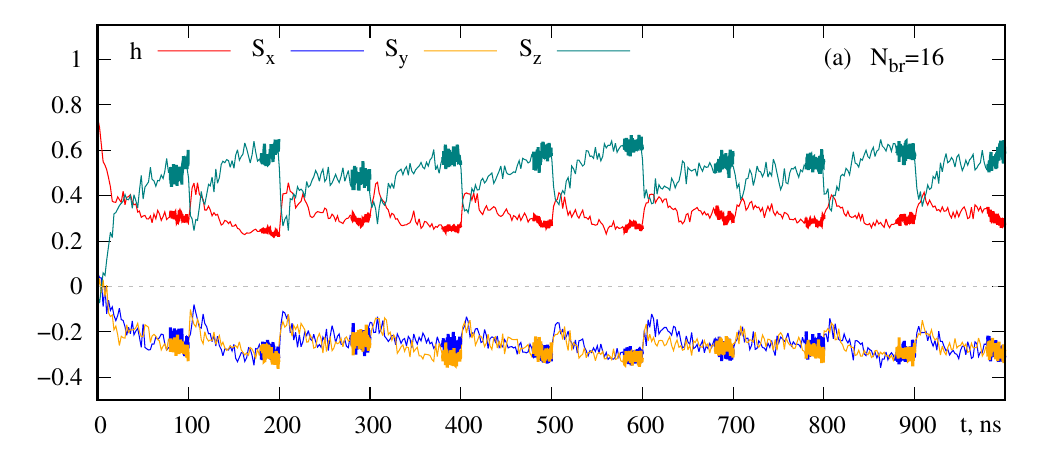}
\includegraphics[clip,width=10cm]{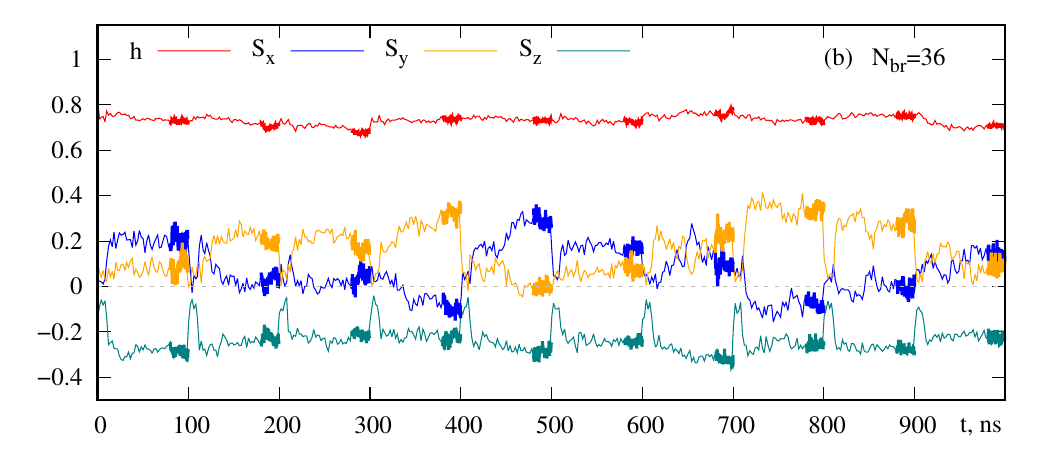}
\caption{\label{BR_vis_evol}(Colour online) The same as in figure~\ref{BR_uv_evol} but for the LCPB under Vis light shown at (a) $N_\text{br}=16$ and (b) $N_\text{br}=36$, unaided self-assembly at $T=490$~K.}
\end{center}
\end{figure}
These observations on the LCPB structure found their quantitative confirmation in the plots for the time evolutions of $h$, $S_x$, $S_y$ and $S_z$, shown in figure~\ref{BR_vis_evol} for two charactreristic cases, (a) $N_\text{br}=16$ and (b) $N_\text{br}=36$. Spontaneously formed collapsed state at $N_\text{br}=16$ is characterized by low brush height $h\approx 0.3$, high order parameter $S_z\approx 0.55$, and both order parameters $S_x$ and $S_y$ being close to $-0.25$, indicating homeotropic anchoring of chromophores. Straightened state at $N_\text{br}=36$ has a high brush height $h\approx 0.7$ and order parameter $S_z$ close to $-0.25$, indicating planar arrangement of chromophores. The order parameters $S_x$ and $S_y$ fluctuate all the time in between $-0.1$ and $0.4$ values, demonstrating a constant drift of nematic director in the $XY$ plane and no evidence for the UPL arrangement. 

\begin{figure}[!t]
\begin{center}
\includegraphics[clip,width=10cm]{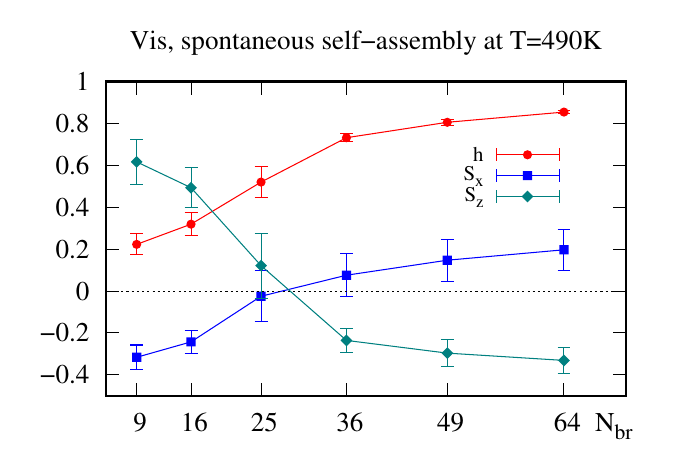}
\caption{\label{BR_Fig_2}(Colour online) The same as in figure~\ref{BR_Fig_1} but for the unaided self-assembly of LCPB under Vis light at $T=490$~K.}
\end{center}
\end{figure}
The values for $h$, $S_z$ and $S_x$, averaged over productive parts of the simulation at various $N_\text{br}$, are shown in figure~\ref{BR_Fig_2}. The transition between collapsed and stretched states of the brush is observed clearly, but it is less sharp comparing to the case of UV light, figure~\ref{BR_Fig_1}, and occurs at higher $N_\text{br}\approx 10-16$. We attribute this effect to stabilization of the collapsed state at $N_\text{br}=4-9$ by both strong attraction between the $t$ chromophores, and wall-induced layering of chromophores. With the increase of $N_\text{br}>16$, the order parameter $S_z$ decreases, dropping down to about $-0.35$ at $N_\text{br}=64$, indicating an enhanced planarity of chromophores in the dense brush regime. Average value of the order parameter $S_x$ (as well as $S_y$, not shown) does not exceed $0.2$, indicating the absence of the UPL state but dynamically changed polydomain LC structure of the chromophore subsystem, seen earlier in the rigth-hand frame of figure~\ref{BR_snap2}. The runs were also conducted at the temperatures $T=480$~K and 500~K with the same result. This indicates that the model LCPB does not self-assemble into the phase with the UPL ordering and instead   produces a globally isotropic planar phase of the LCPB  similar to the one observed under UV light.

This can be explained by the known fact that the LC phases formed by the LC polymers are stabilized by their chemical crosslinking \cite{Symons1993,Symons1999}. Similar effect takes place in our model for the LCPB, where the physical ``crosslinking'' occurs instead. When the initial distribution of chromophore orientations is globally isotropic, then their photoisomerisation into the $t$ type, and the consequent formation of physical links, stabilizes this isotropic distribution. It is evident that to achieve a stable UPL arrangement of chromophores, such an arrangement should be induced by some means first and then it can be fixed by a physically crosslinked LCPB structure.

This forms a basis for the second approach for achieving stable LCPB UPL arrangement considered here, that can be termed as an aided self-assembly. It is based on the experimental studies, where the role of the aligning external field is played by the linearly polarized beam of suitable wavelength
\cite{Uekusa2007,Uekusa2008,Mukai2016,Nagano2018,Li2018}. Assuming that a beam propagates along the $OZ$ axis, its polarization vector $\vhat{p}$ is contained in the $XY$ plane, e.g., along the $OY$ axis. In this case, as a result of cyclic photoisomerization, the $t$ chromophores will be oriented predominantly perpendicularly to $\vhat{p}|| OY$, i.e., they will be confined within the $XZ$ plane. On the other hand, for the case of a dense brush, their orientations are also confined within the $XY$ (see figure~\ref{BR_snap2}). Both factors combined, leaves for the $t$ beads the only option to be aligned predominantly along the $OX$ axis. This effect is modelled in this study by applying external field (\ref{urot}) with the strength of $F=1\textrm{ j}$ and $\vhat{p}$ collinear to the $OY$ axis.

Similarly to spontaneous self-assembly approach, the aided self-assemby is attempted $10$ times at each $N_\text{br}$. Each attempt comprises the 10~$\mathrm{ns}$ reset stage performed at $T=800$~K, followed by the 30~$\mathrm{ns}$ field-aided assembly stage at $T=490$~K, and finally followed by the 60~$\mathrm{ns}$ equilibration stage at the temperature of choice. Last 20~$\mathrm{ns}$ of the relaxation stage of each independent run serve as a productive part of the simulations.

\begin{figure}[!t]
\begin{center}
\includegraphics[clip,width=12cm]{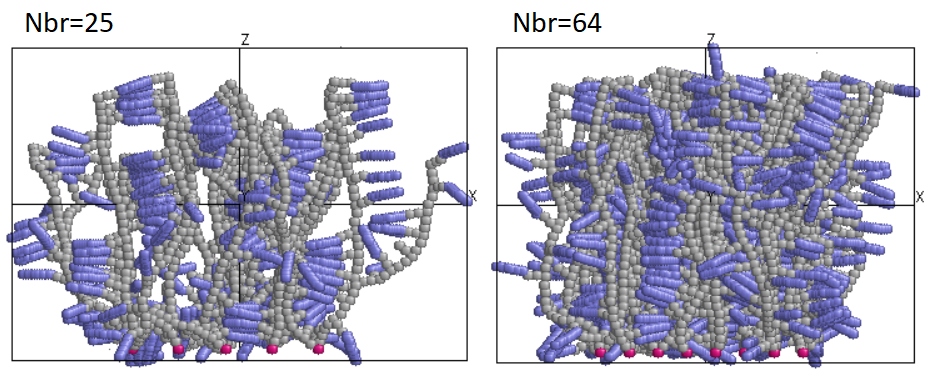}
\caption{\label{BR_snap3}(Colour online) Snapshots showing the UPL arrangement of chromophores, induced initially in LCPB by an external field (\ref{urot}) at $T=490$~K, and then stabilized by physical crosslinks as a result of relaxation at $T=450$~K with no field applied. The cases of $N_\text{br}=25$ and $64$ are shown.}
\end{center}
\end{figure}
As an illustration, we show the snapshots for the LCPB internal structure, obtained as a result of relaxation of the field-induced UPL structure at $T=450$~K for two cases, of $N_\text{br}=25$ and $64$, see, figure~\ref{BR_snap3}. These illustrations clearly show predominantly uniaxial arrangement of chromophores along the $OX$ axis. 

\begin{figure}[!t]
\begin{center}
\includegraphics[clip,width=10cm]{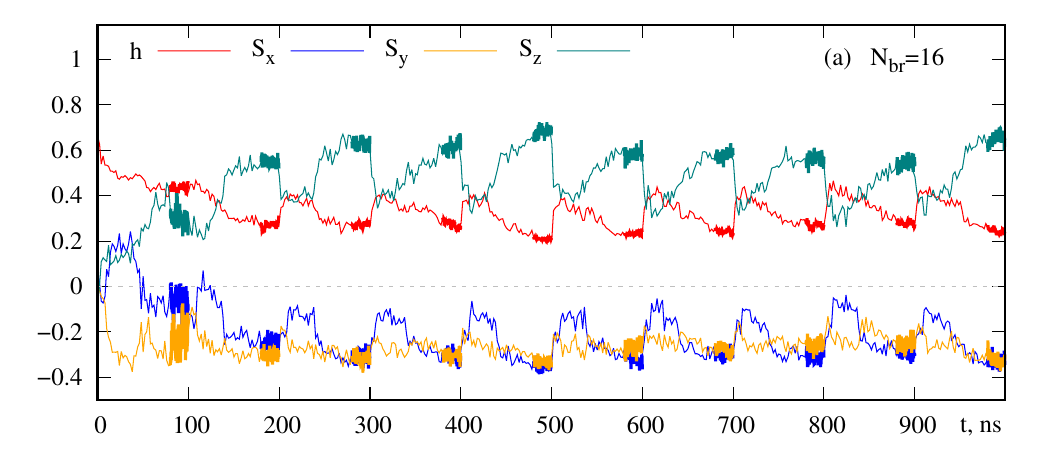}
\includegraphics[clip,width=10cm]{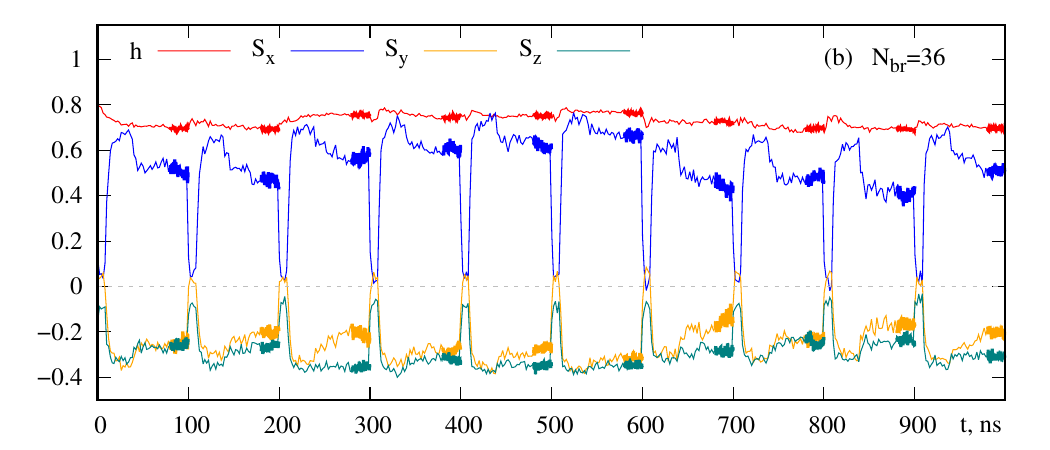}
\caption{\label{BR_vis2_evol}(Colour online) The same as in figure~\ref{BR_vis_evol} but for the case of aided assembly of LCPB and its following relaxation at $T=450$~K.}
\end{center}
\end{figure}
Similarly to the case of unaided self-assembly, we show the plots for the time evolutions of $h$, $S_x$, $S_y$ and $S_z$ for two charactreristic cases, (a) $N_\text{br}=16$ and (b) $N_\text{br}=36$, see figure~\ref{BR_vis2_evol}. The collapsed state (a) is very similar to its counterpart for spontaneous self-assembly, figure~\ref{BR_vis_evol}(a). Straightened state (b), however, is characterized by new features, not present for the case of unaided self-assembly [figure~\ref{BR_vis_evol}(a)]. In particular, during the field-aided assembly stage of each of $10$ runs, both $S_z$ anad $S_y$ drop down sharply, whereas $S_x$ raises up, where the magnitude of all three are essentially non-zero. This indicates a well-defined UPL arrangement of chromophores. It stays stable during the relaxation stage of each run, where $S_z$ and $S_y$ are found to fluctuate around the average values of about $-0.3$, and $S_x$ --- around about~$0.5$. Thus, at $N_\text{br}=36$, the field-aided UPL arrangement of chromophores demonstrates stability being ``fixed'' by the macromolecular network structure of the LCPB.

\begin{figure}[!b]
\begin{center}
\includegraphics[clip,width=10cm]{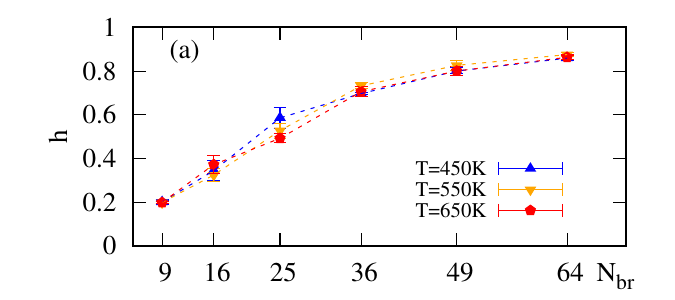}
\includegraphics[clip,width=10cm]{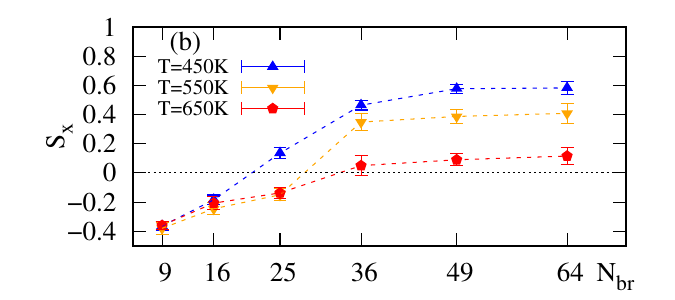}
\includegraphics[clip,width=10cm]{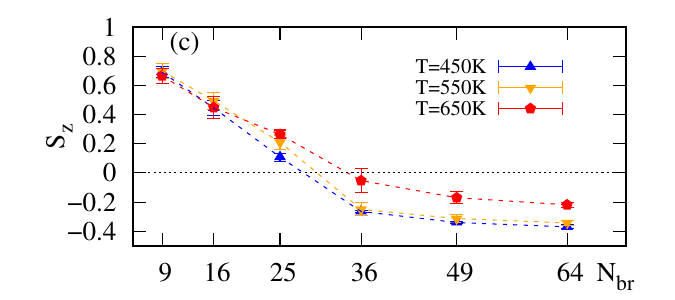}
\caption{\label{BR_Fig_3}(Colour online) (a) Reduced height $h$ and the order parameters (b) $S_z$ and (c) $S_x$ of the chromophores in the LCPB after their UPL arrangement is induced by external field and then relaxed at a given temperature (indicated in the plots) with the field switched off. The data are averaged over $10$ attempts performed at each brush density given by $N_\text{br}$.}
\end{center}
\end{figure}
The set of characteristics, $h$, $S_z$ and $S_x$, averaged over the productive stages of simulations of $10$ independed attempts, are shown in figure~\ref{BR_Fig_3} as functions of the number of molecules $N_\text{br}$, at three temperatures $T=450$~K, $550$~K and $650$~K. We would like to make several observations here. Observation (i) is: the dependence of the reduced height $h$ on $N_\text{br}$ is practically independent of temperature, see figure~\ref{BR_Fig_3}(a). Observation (ii) is: formation of the UPL arrangement  (which can be associated with the case of $S_z< -0.3$ and $S_x>0.4$) is observed at the temperatures $T=450$~K and $550$~K only and is not observed at $T=650$~K, see figure~\ref{BR_Fig_3} (b) and (c). The threshold temperature for the formation of the UPL phase is, therefore, located in between $550$~K and $650$~K. Observation (iii) is: the UPL arrangement is observed only at a certain threshold brush density, $N_\text{br}\geqslant 36$, where the brush is found in the straightened state ($h\approx 0.7$). 

These simulation results prove that three factors are vital for the formation of a stable UPL arrangement of the LCPB: the need for a prealignment of chromophores; the strength of the LC interaction between them, controlled by the temperature; and the strength of the ``memory effect'', provided by a sufficient density of polymer chains. The role of the polymer system is twofold: to induce a required arrangement of its components: backbones are stretched upright  along the $OZ$ axis and chromophores are kept perpendicular to them, i.e., planar with respect to the substrate; and to stabilize the prealigned uniaxial order of chromophores by means of physical crosslinks between polymer chains. All this is possible at the temperatures not much higher than the nematic-isotropic phase transition for the chromophores subsystem. To summarize, the simulations show that to achieve a stable UPL orientation for the LCPB, one requires a brush that is dense enough, and the temperature, where the LC order of chromophores is strong enough. Particular intervals for both characteristics depend on the parameters of the interaction potentials and their practical estimates for real chemical substances require thorough parametrization procedures to be employed.

\section{\label{V}Conclusions}

The development of photo-switchable commanding surfaces, including the possibility of triggering the required anchoring of bulk liquid crystals, has found a lot of experimental interest recently. Computer simulation of such systems faces several difficulties. These are related to the relatively large length and time scales of the problem, and turn problematic  rendering  these effects with the use of chemically-exact models. Instead, one may attempt a more physics-oriented approach, when not the exact chemical details, but rather principal features of the phenomena are taken into account. Such an attempt is made in the current study.

We suggest a mesoscopic model for a liquid crystalline polymer brush, which is built of polymer chains that are grafted to a substrate, each containing chromophoric groups. The model reproduces such principal features of the real systems as: side-chain molecular architecture, semi-rigidity of polymer chains, polarity of chromophoric groups and of a solvent and photoisomerization mimicking that of the azobenzene chromophores. Brush density is defined via the number $N_\text{br}$ of its chains while keeping the dimensions of the simulation box constant. Grafting pattern is regular, in a form of a square lattice, and the cases $N_\text{br}$ from $1$ up to $64$ are considered. We examine the reduced brush height $h$, as well as the orientation order parameters $S_{\alpha}$ of chromophores with respect to respective Cartesian axes $\alpha=x,y,z$. For each parameter set, $10$ independent runs are performed, each started from a high temperature ``reset'' stage which is followed by one or two stages (depending on a simulation type), performed at a required temperature.

First the model brush is considered under ultraviovet light. In this case, all chromophores are assumed to have the properties of the \textit{cis}-isomers of azobenzene, namely: they are polar and non-mesogenic. The brush is this case displays typical properties of the non-liquid crystalline brushes, namely: a transition from a collapsed to a straightened state which takes place already at $N_\text{br}$ as low as $2$. The arrangement of chromophores is weakly planar and isotropic within the plane representing a substrate.

Under a visible light, all chromophores are assumed to represent unpolar mesogenic \textit{trans}-isomers, that are capable of forming a bulk nematic phase at  temperatures lower than the transition temperature $T_{NI}$. Two approaches were attempted aimed at forming the uniaxial planar arrangement of chromophores. In the first one, the brush was given the possibility to self-assemble into the ordered phase at $T<T_{NI}$. No such arrangement was observed at any brush density given by $N_\text{br}$. At low $N_\text{br}$, this was not the case because of the collapsed state of the brush, whereas at higher $N_\text{br}>36$, the arrangement of chromophores is planar but not uniaxial, characterized by a constant rotation of the nematic director within the plane representing a substrate. Typical simulation snapshots show a locally ordered polydomain microstructure of the brush in this case. 

The second approach utilizes prealignment stage for chromophores, performed with the aid of an external field. After this stage, the relaxation stage is performed with the field switched off and the brush being left to relax at a chosen temperature. The results were averaged $10$ independent runs, where the data being averaged was collected during production intervals of the relaxation stages of each run. This made it possible to examine the range of both $N_\text{br}$ and of the temperature $T$, where stable uniaxial planar arrangement can be achieved.

The results obtained prove that three factors are vital for the formation of a stable uniaxial planar arrangement of the brush: the need for a prealignment of chromophores; the strength of the LC interaction between them; and the strength of the ``memory effect'' of the brush. With respect to this, the role of the polymer system is twofold: to induce a required arrangement of its components and to stabilize the prealigned uniaxial order of chromophores by means of physical crosslinks between polymer chains. All this is possible at  temperatures not much higher than $T_{NI}$ and at a sufficient density of the brush given by $N_\text{br}$. The relevant intervals for this model are found in this study, but they  strongly depend on the chosen parameters of the interaction potentials, and their practical estimates for real chemical substances require thorough parametrization procedures to be employed.

The main result of the study is the evidence that the relatively simple mesoscopic model of a complex macromolecular system is capable of describing the possibility to make a commanding surface, which can photo-switch between the orientationally isotropic and uniaxially aligned planar phases. Although, such surfaces have been already developed experimentally, their modelling provides a cheaper and simpler alternative to look for their enhancements or possible changes of their functionality. To this end, we plan to perform further studies involving special parametrizations of the model interaction parameters, aimed at closing the gap between the models of this kind and real chemical architectures.

\section*{Acknowledgements}

The authors acknowledge financial support of the National Academy of Sciences of Ukraine made via the grant program 6541230.
The simulations in this study were performed using facilities of the computing cluster ICMP and the Ukrainian National Grid.
The authors also express their gratitude to the Armed Forces of Ukraine, the National Guard and other relevant law enforcement agencies for their selfless service, 
what made this research work possible.


\ukrainianpart

\title{Рідкокристалічна щітка з фото-перемиканням в якості орієнтуючої поверхні для рідких кристалів: моделювання за допомогою мезоскопічного комп'ютерного моделювання
}
\author[Д. Яремчук, Т. Пацаган, Я. Ільницький]{Д. Яремчук,
Т. Пацаган, Я. Ільницький}
\address{
 Інститут фізики конденсованих систем Національної академії наук України, \\вул. Свєнціцького, 1, 79011 Львів, Україна
}
%
%
%
\makeukrtitle

\begin{abstract}
\tolerance=3000%

Ми розглядаємо мезоскопічну модель рідкокристалічної щітки, яка може служити фотоперемикаючою поверхнею для попередньої орієнтації низькомолекулярних рідких кристалів в об’ємній фазі. Щітка побудована шляхом причеплення до субстрату полімерів із молекулярною архітектурою, що містить бічні ланцюги, причому кожен бічний ланцюг закінчуються хромофорними групами азобензену. При опроміненні ультрафіолетовим світлом хромофори фотоізомеризуються в немезогенний \textit{cis} стан, а вся система перетворюється на звичайну полімерну щітку без орієнтаційного порядку, що може перебувати у двох станах: згорнутому і випрямленому залежно від щільності щеплення. При опроміненні видимим світлом, хромофори фотоізомеризуються в мезогенний \textit{trans} стан, в результаті чого утворюється сітка із ланцюгів за рахунок сильного притягання між хромофорами. Самоскупчення щітки в цих умовах призводить до орієнтаційно ізотропної полідоменної структури. Бажане планарне одновісне впорядкування хромофорів в щітці може бути досягнуте в певних інтервалах температури та щільності причеплення полімерів в результаті двоетапного протоколу приготування. Зовнішній чинник орієнтує хромофори одновісно на першому етапі. На другому етапі система врівноважується при заданій температурі та при вимкненому зовнішньому чиннику. Попередньо орієнтовані хромофори зберігають або втрачають орієнтацію в залежності від величини ефекту пам’яті, властивого сітці ланцюгів, що утворюються на першому етапі, подібно до випадку рідкокристалічних еластомерів, де такі ефекти викликані ковалентними поперечними зв’язками.


\end{abstract}

\end{document}